\documentclass[twocolumn,showpacs,preprintnumbers,amsmath,amssymb]{revtex4}

\usepackage{graphicx}
\usepackage{dcolumn}
\usepackage{bm}
\begin{document}

\preprint{APS/123-QED}

\title{P-Violation Manifested at the Molecular Level - A Simple Means for an Absolute Definition of ``Left" {\em vs.} ``Right"}

\author{Avshalom C. Elitzur}
\affiliation{Unit of Interdisciplinary Studies, Bar-Ilan University,
52900 Ramat-Gan, Israel}

\author{Meir Shinitzky}
\affiliation{Department of Biological Chemistry, Weizmann Institute
of Science, 76100 Rehovot, Israel}

\date{\today}
\begin{abstract}
P-violation of the nuclear weak force was hitherto believed to manifest itself only in asymmetric $\beta$-emission due to neutron decay. Molecules with space asymmetry (like L and D amino acids) are affected by this parity violation, yielding a parity violation energy difference (PVED) between chiral isomers in the order of $10^{-16}$ ev. In specific cases, where an autocatalytic process is combined with interaction with a selective spin isomer of H$_2$O, this tiny PVED can be amplified to a macroscopic detectable level. We describe such a system through which one can transmit a universal definition of ``left" and ``right." The procedure is straightforward, yielding an absolute rather than a statistical definition. The argument is illustrated by paraphrasing Feynman's celebrated thought-experiment of communicating spatial directions to an alien.
\end{abstract}

\maketitle
\begin{flushright}
In memory of Prof. Ofer Lider, \\A rare talent in science and
arts.
\end{flushright}
P-violation, demonstrated first in 1957 \cite{Wu57}, provided for the first time an absolute definition of spatial coordinates. It is believed, however, to be confined to the subnuclear domain, leaving the higher levels (atomic, molecular, etc.) unaffected. In this letter we show that, under specific conditions, P-violation can be detected in the molecular level. We first describe the significance of P-violation with the aid of a {\em gedankenexperiment} due to Feynman. We then propose our simplification of the experiment, based on P-violation at the molecular regime. We conclude with some reflections on the experiment's wider implications.

\section{Invariance and Incommunicability}

Feynman \cite{Fey65} rephrased P invariance in the form of a communication problem. 
Suppose you have established communication with extraterrestrial aliens somewhere 
in the universe and want to describe to them the properties of your world. 
Conceptually, it is possible to send a series of signals conveying some basic 
arithmetical elements, such as prime numbers. Next, some physical measures can be 
communicated, such as the periodic table and some constants of nature. Then, based 
on these measures, you can convey all the properties of your surroundings: sizes, 
colors, chemical properties, etc.

One physical property, however, proves elusive, namely, the spatial directions ``left" and ``right." One can, for example, describe the English script to the aliens, but one cannot tell them what "left-to-right" means. Similarly, when communicating to them some important facts about human biology, such as the heart's position in the human chest, or the right- or left-handedness of chiral molecules, they might as well assume the reverse picture. 

Against this challenge, Feynman proposed utilizing the parity
violation involved in electro-weak nuclear interactions:  ``take a radioactive stuff, a neutron, and look at
the electron which comes from such a beta-decay. If the electron
is going up as it comes out, the direction of its spin is into the
body from the back on the left side" (\cite{Fey65}, p. 103). Obviously, the
experimental demands for such a test are extraordinary
\cite{Wu57}, while the final definition of ``right" {\em vs.}
``left" is given by a quantitative recording.

Based on recent findings, we are now at a stage where we can propose a straightforward and much simpler method of conveying spatial directions, also based, indirectly, on the parity violation induced by the nuclear weak force.

\section{A Chemical Manifestation of P Violation}

The neutral current interaction between the nuclear weak force and
the electron cloud of atoms and molecules is extremely weak and
undetectable. However, in the special case of chiral molecules,
the interaction of the asymmetric electron cloud with the
asymmetric nuclear weak force will in principle be different for
the two enantiomers. One enantiomer will gain $\epsilon_{pv}$ while the other
will lose the same amount of energy i.e., $-\epsilon_{pv}$. The
parity violation energy difference (PVED) will therefore be
PVED=$2\epsilon_{pv}$.

Hegstrom {\em et al.} \cite{Heg80}, as well as Mason and Tranter
\cite{Mas84}, have calculated PVED for chiral carbon molecules,
like amino acids, and found that for such molecules, at around
room temperature, PVED is in the order of $10^{-17}kT$ or
$10^{-17}RT$ per mole. More recent elaborate calculations
\cite{Bak98,Zan98,Mac00} suggested that PVED in such molecules is
actually greater by about one order of magnitude, i.e,
PVED=$10^{-16}kT$. In other words, one enantiomer will be of a
lower energy state of $10^{-17}-–10^{-16}kT$. Translated to
probabilities, in a simple racemic synthesis, the more stable
isomer will amount to an excess of $10^{4}$ to $10^{5}$ molecules
per mole (i. e., $6 \times 10^{23}$). For the specific case of
amino acids, the natural L enantiomers are the more stable ones
\cite{Mas84}.

One might argue that PVED implies that any racemic mixture, i.e., of precisely $50\%$ of each chiral isomer, generally obtained in direct chemical synthesis, may possess optical activity in solution. For example, a solution of DL amino-acids, might be applied for determination of right {\em vs.} left. However, due to the tiny magnitude of PVED, the excess it confers on one of the enantiomers corresponds, even under the most extreme conditions, to optical rotation of less than a few picodegrees. This is much below the maximal hypothetical resolution that any putative polarimeter can achieve. Optical rotations below microdegrees lie within the thermal fluctuations (i.e., ``noise") and are consequently undetectable \cite{Eli94}. Similar considerations apply also for the unique group of  ambi-chiral molecules, like asymmetrical secondary amines, e.g., ethylmethylamine. They  possess a non-zero optical rotation \cite{Mac04}, yet of a similar magnitude to those of ordinary racemic mixtures.

In principle, chiral assemblies associated with interaction with water can adopt an autocatalytic process \cite{Shi04,Sco06}, which will extend the difference between the two enantiomers \cite{Bon91}. Systems of ``chiral enhancement" of this type have been recently described \cite{Kon90,Shi93}. We selected one of them \cite{Shi93} to be used in our set-up for distinction of right {\em vs.} left, described below. It is based on the formation of micellar aggregates in water, a typical autocatalytic process, of lipophilic derivatives of the L and D enantiomers of the amino acid serine. The specific optical rotations of these micelles are strong and in opposite directions. However, the absolute magnitudes of their optical rotations are unexpectedly different by up to $50\%$ \cite{Shi93}. It should be noted that it is assumed that the ambience is such that molecules preserve their 3-dimensional structure for a sufficiently long time. This is especially pertinent to a temperature which should be below $200^\circ-300^\circ$, where processes of racemization or enantiomerization \cite{Avnir} are insignificant. Furthermore, in our selected system such processes, even though negligible, would not alter the final definition of ``right" vs. ``left." 

\section{Simplifying Feynman's Solution}

Back to Feynman's scenario, then, trying to convey the notions of ``right" and ``left" to the aliens, we would instruct them to do the following (see figure):
\begin{enumerate}
\item Synthesize the amino acid serine. Separate its two enantiomers (obviously, you don't know which of them is ``L" and which is ``D" by Earth conventions). Synthesize their N-stearoyl derivatives. Dissolve each of them in water to exactly the same concentration.
\item Establish the following set of Cartesian coordinates, where the Z is along your gravitational vector such that ``down" is towards the ground. The Y-axis is parallel to the ground and is along your detection system (e.g. eyes) and the experiment set-up, such that ``forward" is the direction away from you. It is the X-axis, normal to the ZOY plane, the ``right" and ``left" directions of which we wish to define.
\item Construct the following set-up: An elongated cylindrical tube aligned at the Y axis, divided by a transparent partition, e.g. quartz, sealed by two transparent windows. On each side place a polarizer. One, termed ``the analyzer," is aligned at the Z axis and can be rotated. The other, aligned at the X axis, is fixed. Place a UV light source in front of the X polarizer. On the other side place a light detector behind the analyzer.
\item Place water in both compartments of the tube and adjust the analyzer till no light from the source reaches the detector. Polarizers are thus precisely crossed, i.e., at $90^\circ$.
\item Empty the compartments, dry them and then add to each compartment one or another of the two aqueous solutions of N-stearoyl serine.
\item Now rotate the analyzer slightly ($ \leq 1^\circ $ ). If the light signal is consequently increased, the direction of rotation is ``left" while if it is decreased, the direction is ``right."
\end{enumerate}
\begin{figure}[ht]
\centering\includegraphics[width=8.5cm]{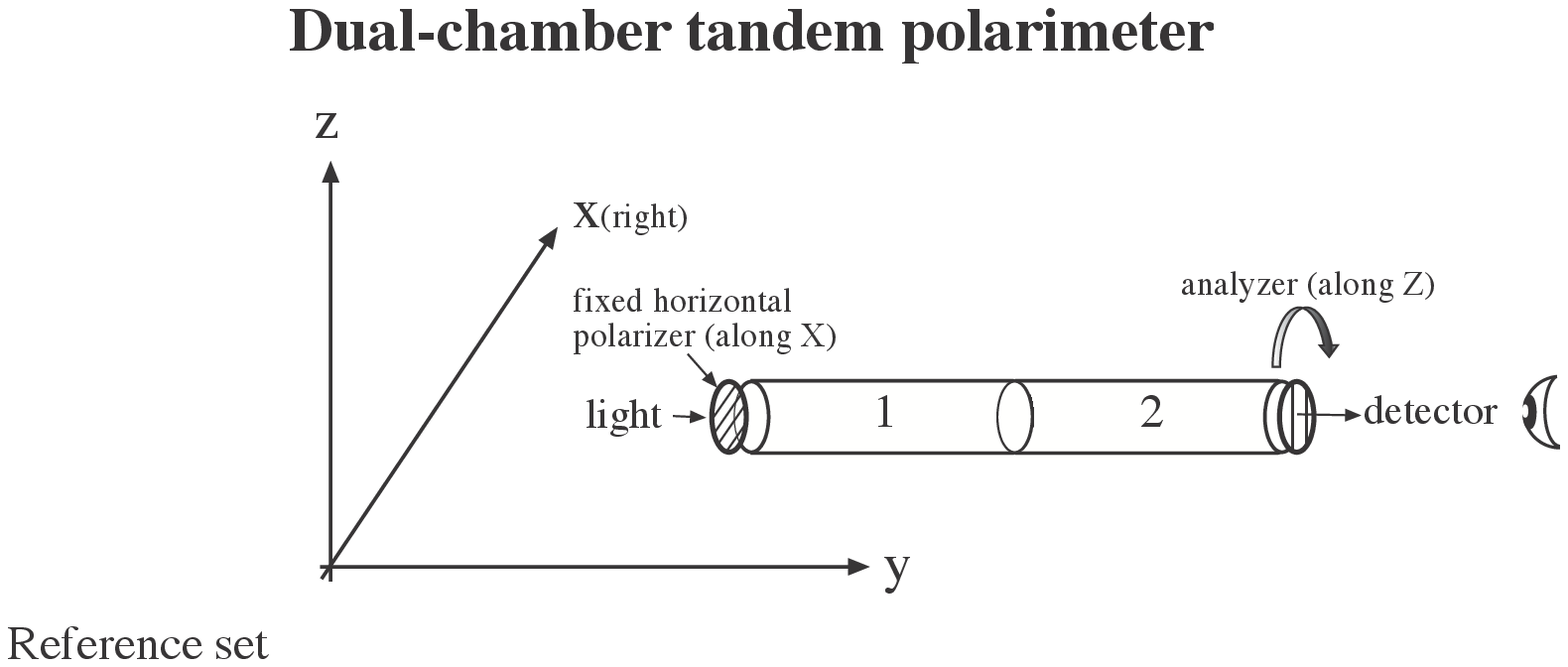}
 \protect\label{fig}
\end{figure}

 \section{Broader Implications}
Molecular P-violation came as a surprise \cite{Schwartz}, 
a reminiscent of the surprise evoked earlier by P-violation itself. 
Its importance obviously extends beyond 
ordinary chemistry, bearing on key issues in theoretical physics. 
Thus Quack \cite{Quack}, in his extensive review (p. 4627), states: ``the role of molecular parity violation for fundamental physics leads us to further motivation for a sustained experimental and theoretical research efforts in this field... A combination of accurate experiments with accurate calculations of the expected effects can be used as a testing ground for the `standard model'."

Finally, it is tempting to speculate that the absolute preference for one enantiomer over the other will turn out to be already inherent to the aliens' own chemical constituents. In other words, biological chemistry based on L amino-acids and D saccharides, etc., might reflect a universal rule rather than an evolutionary coincidence on Earth alone \cite{Quack}.

{\bf Acknowledgments} It is a pleasure to thank Dr. Yosef Scolnik and Prof. David Avnir for inspiring discussions.

\end{document}